\documentstyle[preprint,aps,prl]{revtex}
\tightenlines 
\begin{document}
\draft
\title{Adsorption of a semiflexible polymer onto interfaces and surfaces }
\author{Semjon Stepanow}
\address{Universit\"{a}t Halle, Fachbereich Physik, D-06099 Halle, Germany}
\date{\today }
\maketitle

\begin{abstract}
We consider the adsorption of a semiflexible polymer chain onto interfaces
and surfaces by using the differential equation of the distribution function 
$G(R,L)$ of the end-to-end distance $R$, which is associated with the moment
expansion of the latter. We present the results of the approximative
treatment consisting of taking into account the 2nd and 4th moments in the
differential equation for $G(R,L)$. The essential features of adsorption of
the semiflexible polymer are: {\it i}) the existence of a new local length
scale, which results in two-exponential decay of the monomer density of
adsorbed polymer; {\it ii}) the binding of the semiflexible polymer is
weaker than that for flexible one for both interface and wall. The
approximative theory presented is restricted to the regime of weak
adsorption, where the effect of the rodlike behavior of the polymer on small
scales is weak.
\end{abstract}

\pacs{PACS numbers: 05.20-y, , 36.20-r, 05.40.+j }

\section{Introduction}

Adsorption of polymers onto surfaces and interfaces plays a major role in
polymer adhesion, stabilization of colloidal suspensions, and also in
processes including biological macromolecules such as protein adsorption on
membranes. For many applications (including biopolymers) it is important to
consider the effect of the stiffness on the adsorption. Over many years
there is current interest on adsorption of semiflexible polymers in the
literature \cite{birshtein79}-\cite{kuznetsov/sung} (and citations therein).
Besides the interest in problem of adsorption of semiflexible polymers there
is also large interest in bulk properties of semiflexible polymers \cite
{morse-fredrickson}-\cite{tkachenko-rabin}. Models of semiflexible polymers
have also applications in different topics \cite{liverpool-edwards}- \cite
{burschka-titulaer}.

In the present article we present a theory of adsorption of a semiflexible
polymer generalizing the approach based on the analogy between statistics of
polymer chains and trajectories of a Quantum Mechanical particle at
imaginary time \cite{edwards}. \ According to this analogy the adsorption of
a flexible Gaussian polymer can be mapped onto the problem of binding states
of a quantum mechanical particle in an external potential well \cite
{degennes69}. The key point of our approach consists of taking into account
the higher moments of the end-to-end distribution function of the
semiflexible polymer $G(R,L)$ to derive a differential equation for the
latter, which generalize the Schr\"{o}dinger type equation of the flexible
polymer. In the case of the piecewise constant potential the adsorption
problem can be mapped on a quantum mechanical one: the solutions of the
differential equation for $G(R,L)$ in regions, where the potential is
constant, have to obey the boundary conditions at jump points of the
potential and in infinity. The binding energy and the wave function of the
ground state are obtained as a result.

The article is organized as follows. Section II introduces to the formalism.
Section III and IV consider adsorption onto a symmetric interface and a
surface, respectively. Section V contains our conclusions.

\section{Formalism}

The distribution function of the end-to-end polymer distance of the
Kratky-Porod model \cite{kratky/porod49} is given by path integral as
follows 
\begin{eqnarray}
G({\bf R}-{\bf R}_{0},L) &=&{\cal N}\int D{\bf r}(s)\int_{k}\exp (i{\bf k}(%
{\bf R}-{\bf R}_{0}-\int_{0}^{L}ds\frac{\partial {\bf r}(s)}{\partial s}%
)\prod\limits_{s}\delta ((\frac{\partial {\bf r}(s)}{\partial s}%
)^{2}-1)\times   \nonumber \\
&&\exp (-\frac{l_{p}}{2}\int_{0}^{L}ds(\frac{\partial ^{2}{\bf r}(s)}{%
\partial s^{2}})^{2}),  \label{pint}
\end{eqnarray}
where $l_{p}$ is the persistence length. The product over \ $s$ in Eq.(\ref
{pint}) takes into account the local inextensibility of the polymer chain.
If the polymer interacts with an external field, the potential energy $%
-\int_{0}^{L}V({\bf r}(s))$ should be added in the 2nd exponential of Eq.(%
\ref{pint}). Instead of $G(R,L)$ it is convenient to consider its Fourier
transform $G(k,L)$ 
\begin{equation}
G({\bf k},L)=\int d^{3}R\exp (-i{\bf k}({\bf R}-{\bf R}_{0}))G({\bf R}-{\bf R%
}_{0},L),  \label{fourier}
\end{equation}
which we represent in terms of the moments \footnote{%
Exactly speaking $\mu _{2n}(L)$ are cumulants of the moments.} as follows 
\begin{equation}
G(k,L)=\exp (-\int_{0}^{L}ds\sum_{n=1}^{\infty }\mu _{2n}(s)(k^{2})^{n})
\label{moments}
\end{equation}
Eq.(\ref{moments}) enables one to derive the following differential equation
for $G(R,L)$ 
\begin{equation}
\frac{\partial \ G({\bf R},L)}{\partial L}-\sum_{n=1}^{\infty }(-1)^{n+1}\mu
_{2n}(L)\Delta ^{n}G=0,  \label{sfp3}
\end{equation}
where $\Delta =\nabla ^{2}$ is the Laplace operator. Eq.(\ref{sfp3}) is
exact for a free polymer chain. In the presence of interaction we add the
term $V({\bf R})G({\bf R},L)$ on\ the l.h.s. of Eq.(\ref{sfp3}). In contrast
to the free polymer we cannot now affirm that Eq.(\ref{sfp3}) is still
exact. The reason is that the distribution function of the end-to-end
distance $G(R,L)$ does not possess the Markovian property in $R$ space.
However, it is easy to see that in the formal solution of Eq.(\ref{sfp3})
the potential $V(R)$ appears correctly via a Boltzmann factor. It also
follows from Eq.(\ref{moments}) that as it should be the higher moments $\mu
_{2n}(L)$ suppress the number of conformations contributing to $G(R,L)$.
Thus, Eq.(\ref{sfp3}) (see also (\ref{sfp6})) is expected to take into
account in a correct way the statistical properties of the semiflexible
polymer and its interaction with the external potential.

The 2nd and the 4th moments of the distribution function of the end-to-end
polymer distance of the free polymer chain were computed for the
Kratky-Porod model as \cite{kratky/porod49}-\cite{saitoetal67} 
\begin{equation}
\ \overline{R^{2}}=2l_{p}^{2}(\exp (-\frac{L}{l_{p}})+\frac{L}{l_{p}}-1),
\label{sfp1}
\end{equation}
\begin{equation}
\ \overline{R^{4}}=8l_{p}^{4}(\frac{5L^{2}}{6l_{p}^{2}}-\frac{26L}{9l_{p}}+%
\frac{107}{27}-4\exp (-\frac{L}{l_{p}})-\frac{L}{l_{p}}\exp (-\frac{L}{l_{p}}%
)+\frac{1}{27}\exp (-3\frac{L}{l_{p}})).  \label{sfp2}
\end{equation}
These moments are related for large $L$ according to $\ \overline{R^{4}}=5/3(%
\overline{R^{2}})^{2}$ (for space dimension $d=3$), while for small $L$ the
relation reads$\ \overline{R^{4}}=(\overline{R^{2}})^{2}$. Demanding that $\ 
\overline{R^{2}}$ and $\ \overline{R^{4}}$ given by Eqs.(\ref{sfp1}-\ref
{sfp2}) are identical with those computed from Eq.(\ref{sfp3}) we find $\mu
_{2}(L)$ and $\mu _{4}(L)$ as 
\begin{equation}
\mu _{2}(L)=\frac{l_{p}}{3}(1-\exp (-\frac{L}{l_{p}})),  \label{sfp4}
\end{equation}
\begin{equation}
\mu _{4}(L)=\frac{l_{p}^{2}}{135}(11l_{p}+3l_{p}\exp (-\frac{L}{l_{p}}%
)-24L\exp (-\frac{L}{l_{p}})+l_{p}\exp (-3\frac{L}{l_{p}})-15l_{p}\exp (-2%
\frac{L}{l_{p}})).  \label{sfp5}
\end{equation}
Neglecting all moments in Eq.(\ref{sfp3}) besides the 2nd one gives the
theory of a flexible polymer \cite{edwards}. Notice that Eq.(\ref{sfp3})
with $\mu _{2}(L)$ given by Eq.(\ref{sfp4}) yields the Gaussian distribution
function of the end-to-end distance with the correct 2nd moment given by Eq.(%
\ref{sfp1}). However, Eq.(\ref{sfp4}) with the exact $\mu _{2}(L)$ does not
give anything new on adsorption, because the adsorption problem is related
to the large $L$ behavior of Eq.(\ref{sfp4}). The terms in Eq.(\ref{sfp3})
associated with derivatives $\ \Delta ^{2}G$, $\ \Delta ^{3}G$, ... are
responsible for the deviation of $G(R,L)$ from Gaussian distribution and are
due to the bending energy and the local inextensibility of the polymer.
Notice that despite the presence of the high-order terms in (\ref{sfp3}) the
distribution function $G(R,L)$ associated with Eq.(\ref{sfp3}) tends for
large $L$ to a Gaussian function. The use of Eq.(\ref{sfp3}) in computing
the end-to-end distribution function $G(R,L)$ \cite{freed71}-\cite
{wilhelm/frey96} by taking into account a few moments has the deficiency
that the latter becomes negative for finite $L$ at values of $\ R$, $%
R/L\simeq 1.2$. However, this value of $R/L$ corresponds to the full
stretching of the polymer and is thus outside the practical relevance. Thus,
this deficiency does not immediately concern the adsorption problem, since
in the latter only a few modes are relevant, while the negative part of $%
G(R,L)$ is due to summation over all modes. The first surprise, which
follows from Eq.(\ref{sfp5}), is that $\mu _{4}(L)$ does not vanish for
large $L$, $\mu _{4}(\infty )=l_{p}^{3}11/135$. The nonzero value of $\mu
_{4}(\infty )$ originates from the preasymptotic terms in $\overline{R^{2}}$
and $\overline{R^{4}}$. This property is important for adsorption of a
semiflexible polymer.

To study the adsorption of a semiflexible polymer based on Eq.(\ref{sfp3})
we add on the left-hand side of Eq.(\ref{sfp3}) the term $U(z)\ G$, where $%
U(z)=\chi \theta (-z)-u_{0}\theta (z)\theta (w-z)$ is the potential
associated with an asymmetric interface, and is shown in Fig.1. Adsorption
onto an impenetrable surface can be obtained from $U(z)$ in the limit of
large $\chi $. Integrations in Eq.(\ref{sfp3}) over the transversal
coordinates gives an equation containing only the dependence on $R_{z}\equiv
z$. The three dimensional character of the initial problem remains contained
in the particular values of $\mu _{2}$ and $\mu _{4}$. Thus, the
investigation of adsorption of a semiflexible polymer reduces to the study
of the differential equation 
\begin{equation}
\frac{\partial \ G}{\partial L}-\mu _{2}\partial _{z}^{2}\ G+\mu
_{4}\partial _{z}^{4}\ G+U(z)\ G=0.  \label{sfp6}
\end{equation}
Having in mind to study Eq.(\ref{sfp6}) for large $L$ we have replaced $\mu
_{2}(L)$ and $\mu _{4}(L)$ in Eq.(\ref{sfp6}) by their asymptotic values $%
\mu _{2}(\infty )$ and $\mu _{4}(\infty )$.

In solving Eq.(\ref{sfp6}) we follow the quantum mechanical method of
treating the problem of bound states in a potential well described by a
piecewise constant potential \cite{landau/lifschitzQM}. The solution of Eq.(%
\ref{sfp6}) in the regions A ($z<0$), B ($0\leq z\leq w$), and C ($w<z$),
where the potential $U(z)$ is constant, is chosen in the form $\ G(L,z)=\exp
(-eL)\exp (\pm k\sqrt{\mu _{2}/2\mu _{4}}z)$, where $k$ takes two values $%
k_{1}$ and $k_{2}$ given by 
\begin{equation}
k_{1}=\sqrt{1+\sqrt{-e_{1}}}\,,\ \ \ \ \ \ k_{2}=\sqrt{1-\sqrt{-e_{1}}}\;
\label{sfp7}
\end{equation}
with $e_{1}=-1-\frac{4\mu _{4}}{\mu _{2}^{2}}(e-u)$, and $u$ being the value
of the potential ($\chi $, $-u_{0}$, $0$) in A, B, and C, respectively.
Henceforth we will measure the energy and distances in units of $\mu
_{2}^{2}/4\mu _{4}$ and $\sqrt{2\mu _{4}/\mu _{2}}$, respectively. Notice
that taking into account higher derivatives in Eq.(\ref{sfp6}) will results
in more eigenmodes $k_{i}(-e)$, which contribute to the space modulation of
the wave function $G(L,z)$. The wave vectors $k_{1}$ and $k_{2}$ are real if 
$e_{1}$ is negative and smaller than $-1$. \ We will assume in the following
that the energy obeys the condition, $-e<1$, which defines the regime, which
we interpret as weak adsorption. We chose the solution in A and C as 
\begin{equation}
\psi _{A}(L,z)=\exp (-eL)(a_{1}\exp (k_{a,1}z)+a_{2}\exp (k_{a,2}z)
\label{sfp8}
\end{equation}
\begin{equation}
\psi _{C}(L,z)=\exp (-eL)(c_{1}\exp (-k_{c,1}z)+c_{2}\exp (-k_{c,2}z),
\label{sfp9}
\end{equation}
where $k_{c}=k_{a}$ for a symmetric potential. Expecting that in the region
B the energy obeys the condition, $-e>u_{0}$, we obtain that $k_{1}$ is real
while $k_{2}$ is imaginary, so that the solution in B can be written in the
form 
\begin{equation}
\psi _{B}(L,z)=\exp (-eL)(b_{1}\exp (k_{b,1}z)+b_{2}\exp
(-k_{b,1}z)+b_{3}\cos (k_{b,2}z)+b_{4}\sin (k_{b,2}z)).  \label{sfp10}
\end{equation}
As in the case of a flexible polymer \cite{degennes69} the density profile
of the monomers in the approximation of the ground state dominance is
proportional to $\mid \psi (z)\mid ^{2}$, while the wave function itself
gives the distribution of the chain end, and, thus, should be positive
quantity. In the following we will consider the adsorption of a semiflexible
polymer onto a symmetric potential well, and onto an impenetrable surface,
separately.

\section{Adsorption onto a symmetric interface}

The coefficients $a_{1}$, ..., $c_{2}$\ in Eqs.(\ref{sfp8}-\ref{sfp10}) have
to be computed from the boundary conditions, which consist in continuity of
the function $\psi (z)$ and its three derivatives \ at the boundaries $z=0$
and $z=w$. The boundary conditions yield a linear homogeneous system of
eight equations, which have a nonzero solution, if its determinant $\Delta
_{ac}$ is zero. The condition $\Delta _{ac}=0$ gives the energy eigenvalues.
In the vicinity of the localization transition (small $u_{0}$) the energy
eigenvalue and the amplitude in Eqs.(\ref{sfp8}-\ref{sfp10}) are obtained
respectively as 
\begin{equation}
-e=\frac{w^{2}u_{0}^{2}}{8}(1-\frac{\sqrt{2}}{8}wu_{0}+...),  \label{sfp11}
\end{equation}
$a_{1}=-\ (u_{0}/8)(1-\exp (-\sqrt{2}w))$, $a_{2}=1$, $b_{1}=(u_{0}/8)\exp (-%
\sqrt{2}w)$, $\ b_{2}=u_{0}/8$, $b_{3}=1$, $b_{4}=(\sqrt{2}/4)\sqrt{u_{0}}\
w\ $, $c_{1}=(u_{0}/8)(1-\exp (\sqrt{2}w))$, $c_{2}=1$. The prefactor in
front of the wave function is determined from the normalization of the
latter. The leading term in Eq.(\ref{sfp11}) coincides exactly with the
energy obtained for a flexible polymer after using the same units (the
prefactor $1/8$ will become $1/4$, if lengths are measured in units of $\mu
_{2}$). It follows from Eq.(\ref{sfp11}) that the energy level for the
semiflexible polymer is higher in comparison to that of the flexible
polymer, so that the binding of the semiflexible polymer is weaker in
comparison to the flexible one. This result can be interpreted as follows.
Let us consider a small piece of the polymer with number of segments $n$,
which is in contact with the interface. The size of that piece is
proportional to $n$ for a semiflexible polymer and to $\sqrt{n}$ for a
flexible one. Thus, the number of contacts of the flexible polymer with the
interface is larger, which results in a larger energy win. The consequence
of the smaller binding energy of the semiflexible polymer is that the space
distribution of the semiflexible polymer is broader in comparison to that of
the flexible polymer. This finding is in agreement with the results of van
Eijk and Leermakers \cite{vaneijk-leermakers}.

However, the most striking difference in comparison to adsorption of a
flexible polymer is the appearance of the 2nd length scale $\xi _{sf}\sim
1/k_{1}$, which at the localization transition is of order of magnitude of
the persistence length $l_{p}$. This length is due to the fourth derivative
in Eq.(\ref{sfp3}). It is finite at the localization transition, where the
localization length $\xi \sim 1/k_{2}$ is large.

Notice that the amplitudes $b_{1}$, $b_{2}$, $a_{1}$, and $c_{1}$ vanish at
the localization transition. The local length $\xi _{sf}$ increases with
increase of the depth of the potential well $u_{0}$ and converges towards
the localization length $\xi $, which decreases with increase of $u_{0}$.
The increase of $\xi _{sf}$ means that the fraction of monomers associated
with local order increases. The both lengths $\xi _{sf}$ and $\xi $ approach
each other, if $e$ approaches the value $-1$. In increasing the depth of the
potential well, the energy tends to the value $-1$, while the wavenumbers $%
k_{1}$ and $k_{2}$ approach the value $1$. It appears that at $-e=1$ the
wave function becomes exactly zero. The threshold value $u_{crit}$
corresponding to $-e=1$ is obtained for $w=1$ as $u_{sa}=4.24$. Such a state
does not exist, so that the description of adsorption by taking into account
the fourth moment in Eq.(\ref{sfp6}) breaks down at $-e\geq 1$. For $%
u>u_{sa} $ the rodlike character of the polymer chain is expected to play a
dominant role. The higher moments in Eq.(\ref{sfp6}) have to be taken into
account to describe the adsorption for $u>$ $u_{sa}$.

\section{Adsorption onto a surface}

The piecewise constant potential, which models the adsorption onto a wall,
is obtained from that shown in Fig.1 in the limit $\chi \rightarrow \infty $%
. To ensure that the wave function is zero at $z=0$ we take the latter in
the region B as 
\begin{equation}
\psi _{B}(L,z)=\exp (-eL)(b_{1}\sinh (k_{b,1}z)+b_{2}\sin (k_{b,2}z)).
\label{sfp13}
\end{equation}
The wave function in the region C is chosen according to Eq.(\ref{sfp9}).
The coefficients $b_{1}$, $b_{2}$, and $c_{1}$ are expressed through the
coefficient $c_{2}$ by using three boundary conditions at $z=w$. $c_{2}$ can
be fixed due to normalization of $\psi (z)$. The fourth boundary condition
gives the energy eigenvalue, which is restricted to the condition $e<-1$.
The threshold value for the depth of the potential well is obtained for $w=1$
as $u_{t}$ $=7.8$. The amplitudes $a_{1}$, ..., $c_{1}$ are obtained in the
vicinity of the threshold as: $b_{1}=0.03$, $b_{2}=0.6$, $c_{1}=-1.2$, $%
c_{2}=1$. Notice that the amplitude $c_{1}$ controlling the decay of the
part of the wave function, which is due to the semiflexible nature of the
polymer, differs from zero at the localization transition.

Let us compare the adsorption of the semiflexible polymer with that of a
flexible chain. The threshold value $u_{t,fl}$ for adsorption of a flexible
polymer onto a surface is obtained as $4.94$ in units used for the
semiflexible polymer ($w=1$), and is thus smaller than the value $u_{t}=$ $%
7.8$ obtained for the semiflexible polymer. We have obtained that both
thresholds approach each other with increasing the width of the potential
well $w$. The comparison of the monomer density as function of the distance
to the surface for flexible and semiflexible polymer is plotted in Fig.2.
Notice that the lengths and the energy are given in units stated in the text
after Eq.(10). The distribution of the ends of the polymer, which is given
by the wave function itself, behave qualitatively in the same way. Notice
that the monomer density of the semiflexible polymer in the vicinity of the
wall is lower than that for flexible polymer. This explains the higher value
of the threshold for the semi-flexible polymer. To understand this
qualitatively let us consider the effect of the wall on an ideal polymer
coil, which is brought in a weak contact with the wall. It is intuitively
clear that the deformation of the semiflexible coil demands higher external
force, or equivalently the repulsion force of the wall is higher for
semiflexible coil. The lower density of the monomers as a function of the
distance to the wall for small distances is the consequence of this
circumstance. Notice that our prediction is in disagreement with MD
simulations \cite{kramarenkoetal96}, which predict that the semiflexible
polymer adsorbs easily. The adsorption state was studied in \cite
{kramarenkoetal96} under the condition that one end of the polymer, which
has finite number of segments, is fixed at the surface. The adsorption
threshold obtained in this way corresponds to the case, when the loop length
becomes comparable with the contour length of the polymer under
investigation. At the localization transition threshold we obtained, $%
u=u_{t} $, the loop length is infinite. The adsorbed state studied in \cite
{kramarenkoetal96} is expected to correspond to the regime of strong
adsorption, where an opposite behavior as we predicted is expected (see \cite
{vaneijk-leermakers} for a related discussion of adsorption onto an
interface).

As in the case of adsorption onto a symmetric interface, the wave function
becomes zero at $-e=1$, which imposes a restriction on the applicability of
the present theory, which thus is restricted to the regime of weak
adsorption. The value $-e=1$ corresponds to the depth of the potential well $%
u_{sa,w}=15.5$ for $w=1$. The effect of the rodlike character of the
localized polymer is expected to be strong for $u>u_{sa,w}$. This is in
agreement with the prediction of a liquid-crystalline phase made by
Kuznetsov and Sung \cite{kuznetsov/sung}. To describe the strong adsorption (%
$u>u_{sa,w}$), the higher derivatives in Eq.(\ref{sfp6}) have to be taken
into account.

\section{Conclusion}

To summarize, we have considered adsorption of a semiflexible polymer chain
onto interfaces and surfaces by using the differential equation for the
end-to-end distribution function of the polymer chain associated with the
moment expansion of the latter. The present study shows that the adsorption
of the semiflexible polymer is qualitatively different from that of a
flexible one. The essential features of adsorption of the semiflexible
polymer are: {\it (i)} the existence of a new local length scale, which
results in two-exponential decay of the monomer density of adsorbed polymer;
({\it ii}) the binding energy of the semiflexible polymer is weaker than
that for flexible polymer for both interface and wall. The approximative
theory presented here is restricted to the regime of weak adsorption ($%
u<u_{sa,w}$), where the effect of the rodlike behavior of the polymer on
small scales is weak.

\acknowledgments 
I acknowledge a support from the Deutsche Forschungsgemeinschaft (DFG),
grants: Ste 981/1-1 and SFB 418. I acknowledge stimulating discussions with
H. Orland, T. Garel, and J-U. Sommer.

\bigskip

Fig.1 The interaction potential with the interface and the wall.

Fig.2 The monomer distribution $\Psi (z)^{2}$ of adsorbed flexible (dotted
line) and semi-flexible polymer at a surface. $u_{0}=9$, $w=1$.


\begin{references}
\bibitem{birshtein79}  T. M. Birshtein, E. B. Zhulina, and A. M. Skvortsov,
Biopolymers {\bf 18}, 1171 (1979); E. B. Zhulina, T. M. Birshtein, and A. M.
Skvortsov, Biopolymers {\bf 19}, 805 (1980).

\bibitem{maggs89}  A. C. Maggs, D. A. Huse, and S. Leibler, Europhys. Lett. 
{\bf 8}, 615 (1989).

\bibitem{gompper/burkhardt89}  G. Gompper and T. W. Burkhardt, Phys. Rev. A. 
{\bf 40}, R6124 (1989).

\bibitem{yethirajetal}  A. Yethiraj, S. K. Kumar, A. Hariharan, and K. S.
Schweizer, J. Chem. Phys. {\bf 100}, 4691 (1994).

\bibitem{kumaretal}  S. K. Kumar, A. Yethiraj, K. S. Schweizer, and F. A. M.
Leermakers, J. Chem. Phys. {\bf 103}, 10332 (1995).

\bibitem{wuetal}  D. T. Wu, G. H. Fredrickson, and J. P. Carton, J. Chem.
Phys. {\bf 104}, 6387 (1996).

\bibitem{kramarenkoetal96}  E. Y. Kramarenko, R. G. Winkler, P. G. Khalatur,
A. R. Khokhlov, and P. Reineker, J. Chem. Phys. {\bf 104, }4806 (1996).

\bibitem{lindenetal96}  C. C. Linden, F. A. M. Leermakers, and G. J. Fleer,
Macromolecules {\bf 29}, 1172 (1996).

\bibitem{vaneijk-leermakers}  M. C. P. van Eijk and F. A. M. Leermakers, J.
Chem. Phys. {\bf 109}, 4592 (1998).

\bibitem{kuznetsov/sung}  D. V. Kuznetsov and W. Sung, J. Phys. II France 
{\bf 7}, 1287 (1997); J. Chem. Phys. {\bf 107}, 4729 (1997); Macromolecules 
{\bf 31}, 2679 (1998).

\bibitem{morse-fredrickson}  D. C. Morse and G. H. Fredrickson, Phys. Rev.
Lett. 73, 3235 (1994).

\bibitem{gupta-edwards}  A. M. Gupta and S. F. Edwards, J. Chem. Phys. {\bf %
98}, 1993 (1558).

\bibitem{tkachenko-rabin}  A. Tkachenko and Y. Rabin, Macromolecules {\bf 28}%
, 8646 (1995).

\bibitem{liverpool-edwards}  T. B. Liverpool and S. F. Edwards, Phys. Rev.
Lett. {\bf 75}, 3016 (1995); T. B. Liverpool, R. C. Ball, and S. F. Edwards,
Europhys. Lett. {\bf 30}, 181 (1995); T. B. Liverpool and S. F. Edwards, J.
Chem. Phys. 103, ... (1995).

\bibitem{burschka-titulaer}  M. A. Burschka and U. M. Titulaer, J. Stat.
Phys. {\bf 25}, 569 (1981).

\bibitem{edwards}  S. F. Edwards, Proc. Phys. Soc. {\bf 85}, 613 (1965).

\bibitem{degennes69}  P. G. de Gennes, Rep. Prog. Phys. {\bf 32}, 187 (1969).

\bibitem{kratky/porod49}  O. Kratky, and G. Porod, Recl. Trav. Chim.
Pays-Bas {\bf 68}, 1106 (1949).

\bibitem{hermans/ullman52}  J. J. Hermans and R. Ullman, Physica {\bf 18},
951 (1952).

\bibitem{heineetal62}  S. Heine, O. Kratky, and J. Roppert, Makromol. Chem. 
{\bf 56}, 150 (1962).

\bibitem{saitoetal67}  N. Saito, K. Takahashi, and Y. Yunoki, J. Phys. Soc.
Jpn. {\bf 22}, 219 (1967).

\bibitem{freed71}  K. F. Freed, J. Chem. Phys. {\bf 54}, 1453 (1971).

\bibitem{wilhelm/frey96}  J. Wilhelm and E. Frey, Phys. Rev. Lett. {\bf 77},
2581 (1996).

\bibitem{landau/lifschitzQM}  L. D. Landau and E. M. Lifschitz, Quantum
Mechanics (Nauka, Moscow, 1974); \S 22, 2nd problem. \newpage Figure captions
\end{references}
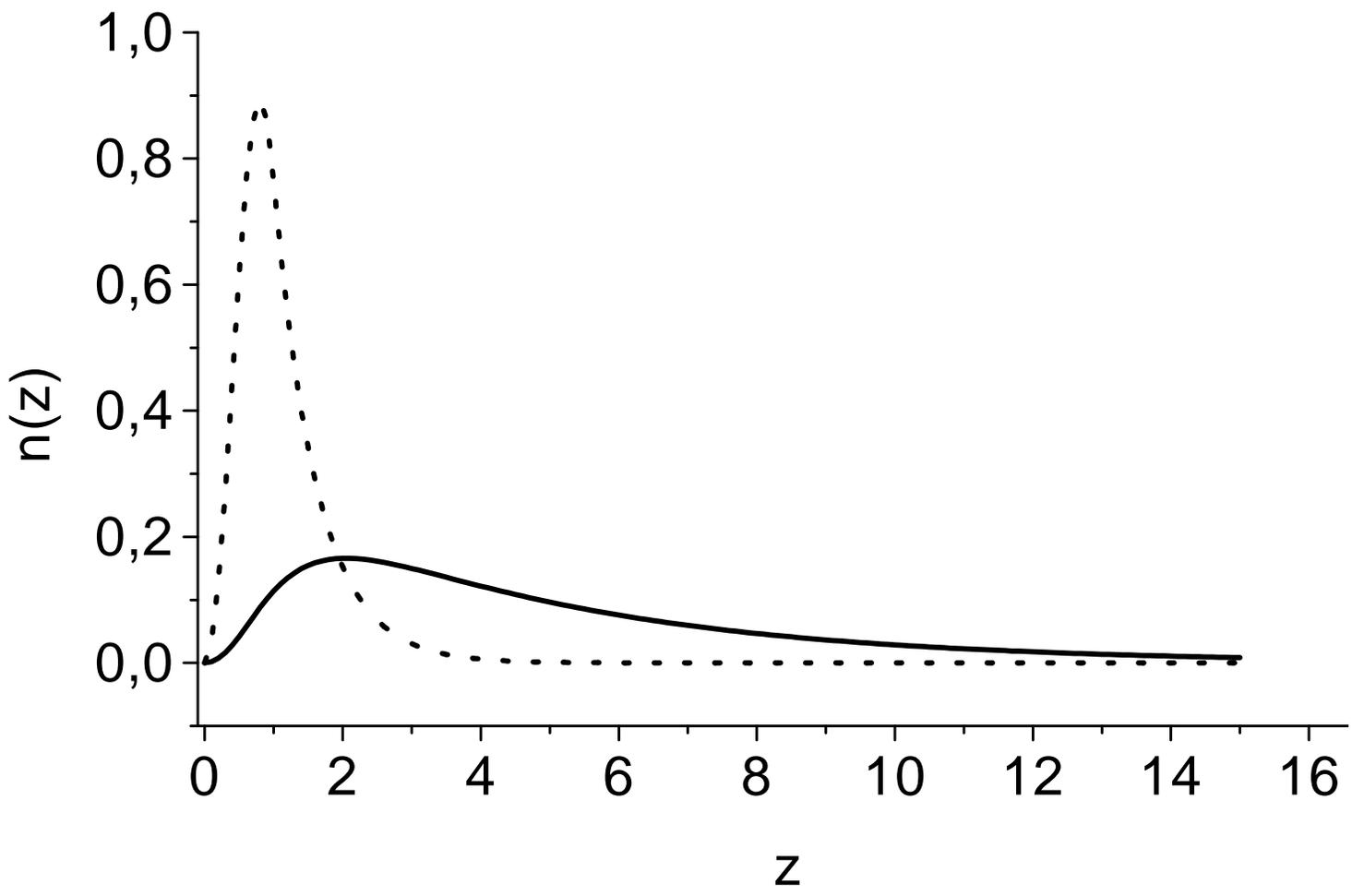
\end{document}